\documentclass{aip-cp}
\usepackage[numbers]{natbib}

\newcommand{\De}{\Delta}

\newcommand{\eps}{\varepsilon}

\begin{document}

\title{Open Questions On Nuclear Collective Motion}

\author[aff1]{S. Frauendorf\corref{cor1}}
\affil[aff1]{Department of Physics, University Notre Dame, IN 46557, USA }
\corresp[cor1]{sfrauend@nd.edu}

\maketitle

\begin{abstract}
The status of the macroscopic and microscopic description of the collective quadrupole modes is reviewed, where  limits due to non-adiabaticity and decoherence are exposed.
The microscopic description of the yrast states in vibrator-like nuclei in the framework of the rotating mean field is  presented.
\end{abstract}

\section{INTRODUCTION}
\begin{figure}[h]
\centerline{\includegraphics[width=6.6cm]{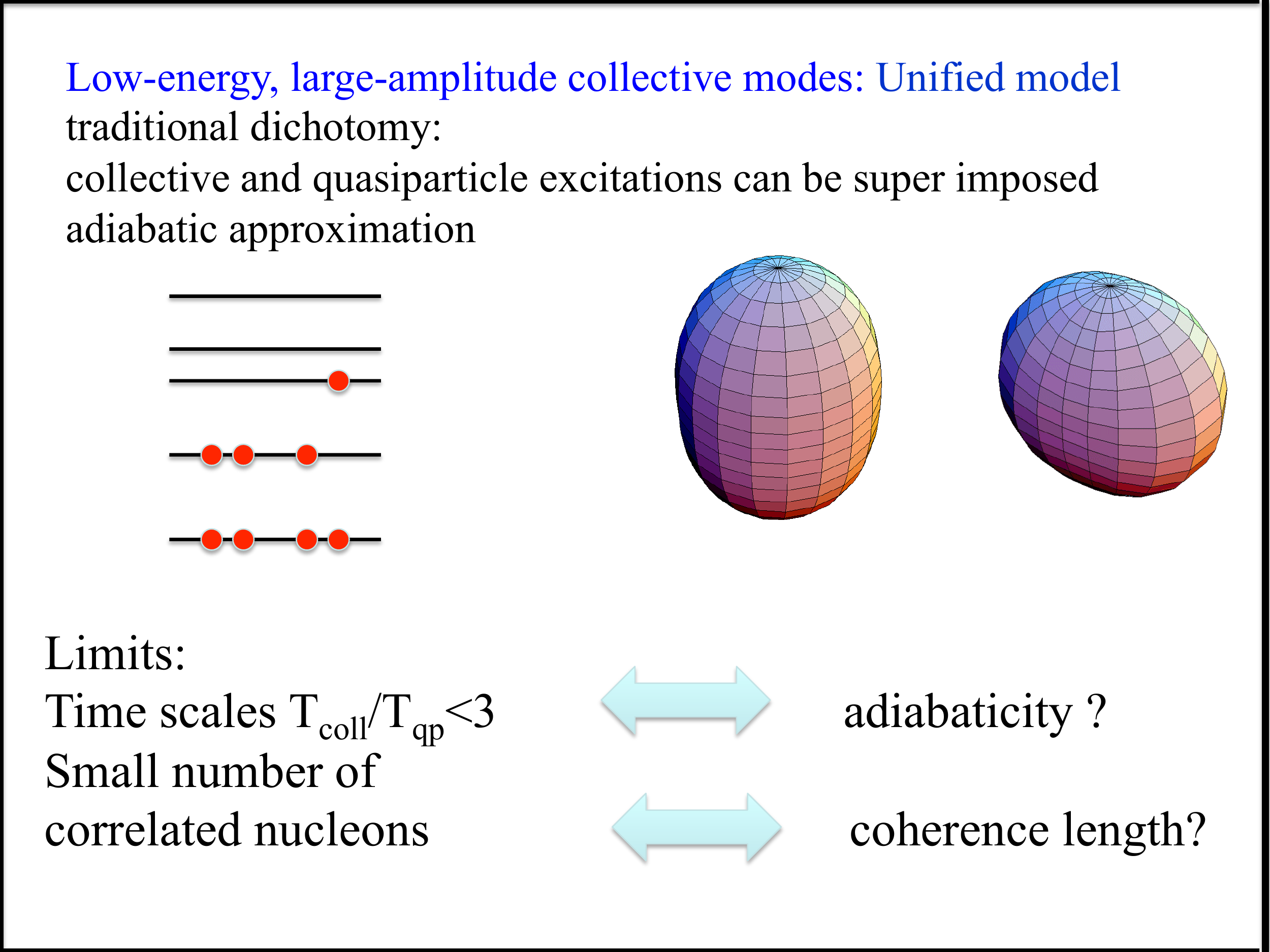}\hspace*{0.5cm}\includegraphics[width=6cm]{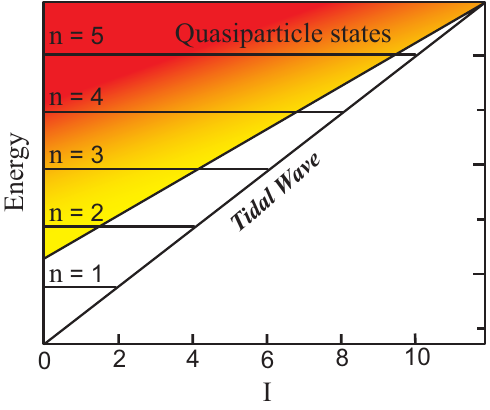}}
\caption{\label{fig:decoherence} Left: The limits of the Unified Model. Right: Schematic representation of the location 
of the collective quadrupole vibrational excitations relative to the quasiparticle excitations. 
The darker shades  indicate higher densities of quasiparticle states. From Ref. \cite{Frauendorf15}.}
\end{figure}

The year 2016 celebrates  the 40$^{th}$  Anniversary of the Nobel Prize for
A. Bohr, B. Mottelson and L. Rainwater, which was awarded for their discovery that nuclei may
have  a non-spherical shape. Bohr and Mottelson casted this innovative concept into the Unified Model (UM),
which allows us to classify the low-lying states of open shell nuclei. Their monograph Nuclear Structure Vol. II: Nuclear
Deformations \cite{BMII} exposes this invaluable tool in great detail. The UM bases on the dichotomy of the  "collective" degrees of freedom, which
describe the shape of the nucleus, and the  "intrinsic" degrees of freedom, which are particle-hole configurations or quasiparticle configurations.
Any number of bosonic collective excitations can be superimposed on the fermionic intrinsic states.  Clearly this is an idealization. 
Nuclei are composed of a relative small number number of nucleons 
compared to other many-body systems. As a consequence, the "granular structure" of the collective degrees of freedom appears already  after 
the excitation of few quanta, which results in a progressive decoherence of the collective modes. The right side of
Fig. \ref{fig:decoherence} illustrates the point in a schematic way for a vibrational nucleus: 
The multi-phonon excitations move very soon into the region of the quasiparticle excitations to which they couple. The collectively enhanced E2 transitions
between the  members of the vibrational multiplets will fragment over the increasingly dense background quasiparticle excitations.
 The collectively enhanced transitions are restricted to the white adiabatic region of the figure, i. e. to the one- and two-phonon 
states and   to  the yrast region for larger spin.    

The purely collective models assume adiabaticity of the collective motion explicitly or implicitly.
Their realm is restricted to the  white adiabatic area in Fig. \ref{fig:decoherence}. 
Two models are widely used to describe the collective excitations of even-even nuclei: the Bohr Hamiltonian (BH) and the Interacting Boson Model (IBM). 
  I restrict to the quadrupole mode,
although both models have also been  been applied to the octupole mode.
  I will briefly review the status of
both  models. Outside, the coupling between the quasiparticle and collective degrees
of freedom must be treated in a non-perturbative way.  There are two different aspects to be taken into consideration: non-adiabaticity and decoherence. 
Near yrast line the level density remains low, and the study of individual quantal states is appropriate.   
Their structure can be well understood as interplay between collective 
and single particle degrees of freedom. The collective and the 
nucleonic motion are strongly coupled, because the time scales of the collective and the single particle 
motion are the same.  The collective motion, which is of rotational type,  can be treated 
in the framework of mean-field theory. Tidal Waves in weakly deformed nuclei are new phenomenon of this regime, which I will  
discuss  in this context.  
Finally I will address the question to what extend the nucleonic underpinning supports the collective degrees of freedom. 
 The question arises when approaching shell closures or when
 moving  away from the yrast line, where the collective modes fragment among the rapidly increasing number of quasiparticle excitations.
 The concept of coherence length is introduced to quantify the resolution limit of the collective degrees of freedom.
 Part of the presented  material is published in the recent review \cite{Frauendorf15}.

\section{The Bohr Hamiltonian}\label{sec:BH}

The collective states are represented by wave functions of the components 
$\alpha_\mu$ ($\mu=-2$, $\ldots$, $2$) of the scale-free quadrupole deformation tensor 
which are expressed by the five-dimensional spherical polar coordinates 
\begin{equation}\label{eqn-bohr-q}
\alpha_\mu = \beta \left[ \cos\gamma\, {\cal D}^{(2)}_{0,M}(\Omega) +
\frac{1}{\sqrt{2}} \sin\gamma \left[ {\cal D}^{(2)}_{2,M}(\Omega) +
{\cal D}^{(2)}_{-2,M}(\Omega) \right] \right],
\end{equation}
where $\Omega$ are the Euler angles specifying the orientation of the shape.

\subsection{The Geometric Collective Model}\label{sec:GCMphen}
The Geometric Collective Model (GCM) is a parametrized version of the BH based on an expansion into scalars 
of increasing power constructed from the coordinate $\alpha_\mu$ and their conjugate momenta $\pi_\mu$.  Only the quadratic term in $\pi_\mu$ is kept.
 In terms of the
quadrupole deformation variables $\beta$ and $\gamma$ and Euler angles $\Omega$, 
the Bohr Hamiltonian is given by
\begin{equation}
\label{eqn-HGCM2}
H_{GCM2}=\frac{\hbar^2}{\sqrt{5}B_2}\biggl[
T_{\beta\beta}+
\frac{\hat \Lambda^2}{\beta^2}\biggr]+V(\beta,\gamma),
\end{equation}
where
\begin{equation}\label{eqn-Tbb}
T_{\beta\beta}=-\frac{1}{\beta^4}\frac{\partial}{\partial\beta}\beta^4\frac{\partial}{\partial\beta},~~~
\hat\Lambda^2=-\biggl(
\frac{1}{\sin 3\gamma} 
\frac{\partial}{\partial \gamma} \sin 3\gamma \frac{\partial}{\partial \gamma}
- \frac{1}{4}
\sum_{i=1,2,3} \frac{\hat{L}_i^{\prime2}}{\sin^2(\gamma -\frac{2}{3} \pi i)}
\biggr).
\end{equation}

\begin{figure}[t]
\centerline{\includegraphics[width=0.5\linewidth]{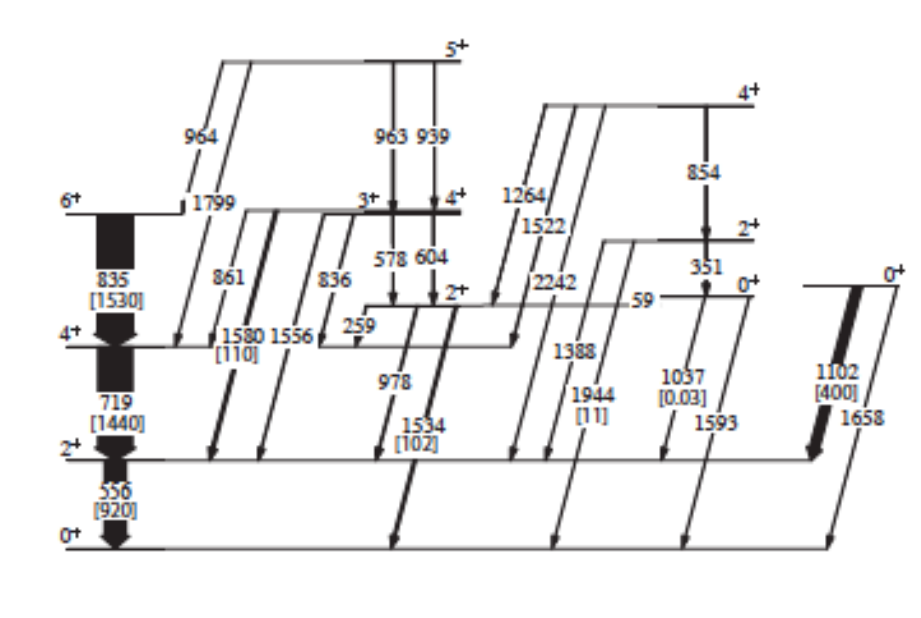}\includegraphics[width=0.5\linewidth]{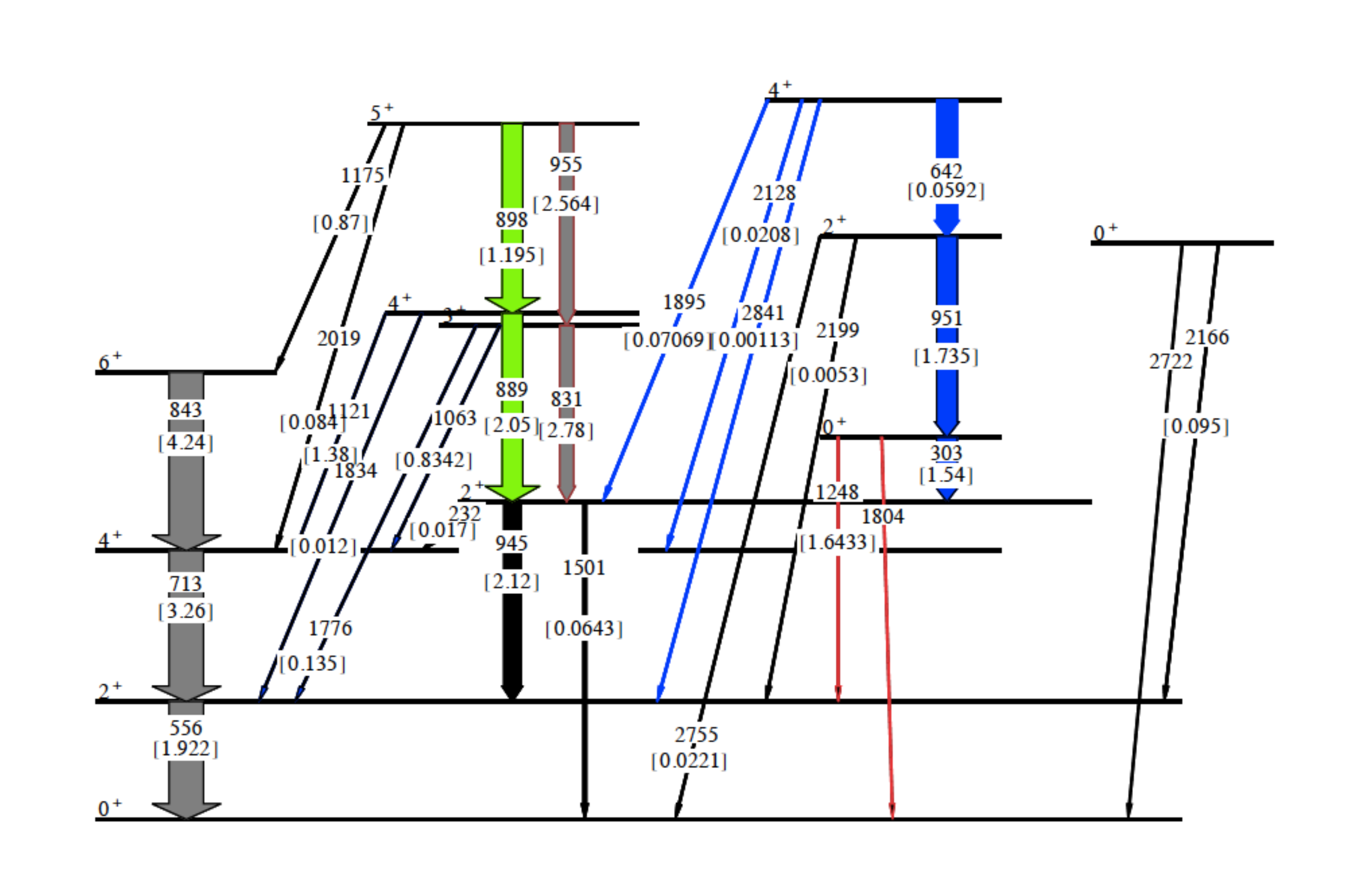}}
\caption{\label{fig:102PdACM} Left: Experimental level scheme and $B(E2)$ strengths  for $^{102}$Pd. 
The number in parenthesis under the transition energies (keV) are the $B(E2)$ values ($e^2$ fm$^4$) for the transitions. 
Data from \cite{Zamfir02}. Right: GCM predictions\cite{Zamfir02} of the energies and $B(E2)$ strengths for $^{102}$Pd  
 normalized to the experimental $E(2^+_1)$ and and $B(E2;2^+_1\rightarrow0^+_1)$
values. The number in parenthesis under the transition energies (keV) are the $B(E2)$ values ($e^2$ fm$^4$) for the transitions. From Ref. \cite{Frauendorf15}
}
\end{figure}

The properties of the experimental  $2^+_1$, $4^+_1$, $6^+_1$, $2^+_2$, $3^+_1$, $4^+_2$, $0^+_2$, $4^+_3$ states can be classified
by assuming that the collective potential contains only three terms,
\begin{equation}\label{eqn-VGCM2}
V_{A}(\beta,\gamma)=\frac{1}{\sqrt{5}}C_2\beta^2-\sqrt{\frac{2}{35}}C_3\beta^3\cos3\gamma+\frac{1}{5}C_4\beta^4.
\end{equation}   
Caprio \cite{Caprio2par} demonstrated that the structure of the collective wave function is determined by  two dimensionless parameters, 
which provides a more complete scheme than the traditional classification into rotational and vibrational nuclei.
He provides  figures of the ratios $E(I)/E(2^+_1)$ and $B(E2;I\rightarrow I')/B(E2;2^+_1\rightarrow0^+_1)$, which allows one to extract the 
 parameters, and discusses fitting strategies. Fig. \ref{fig:102PdACM} shows $^{102}$Pd  which exemplifies the general quality of the GCM phenomenology. 

\begin{figure}[h]
\centerline{\includegraphics[width=0.5\linewidth]{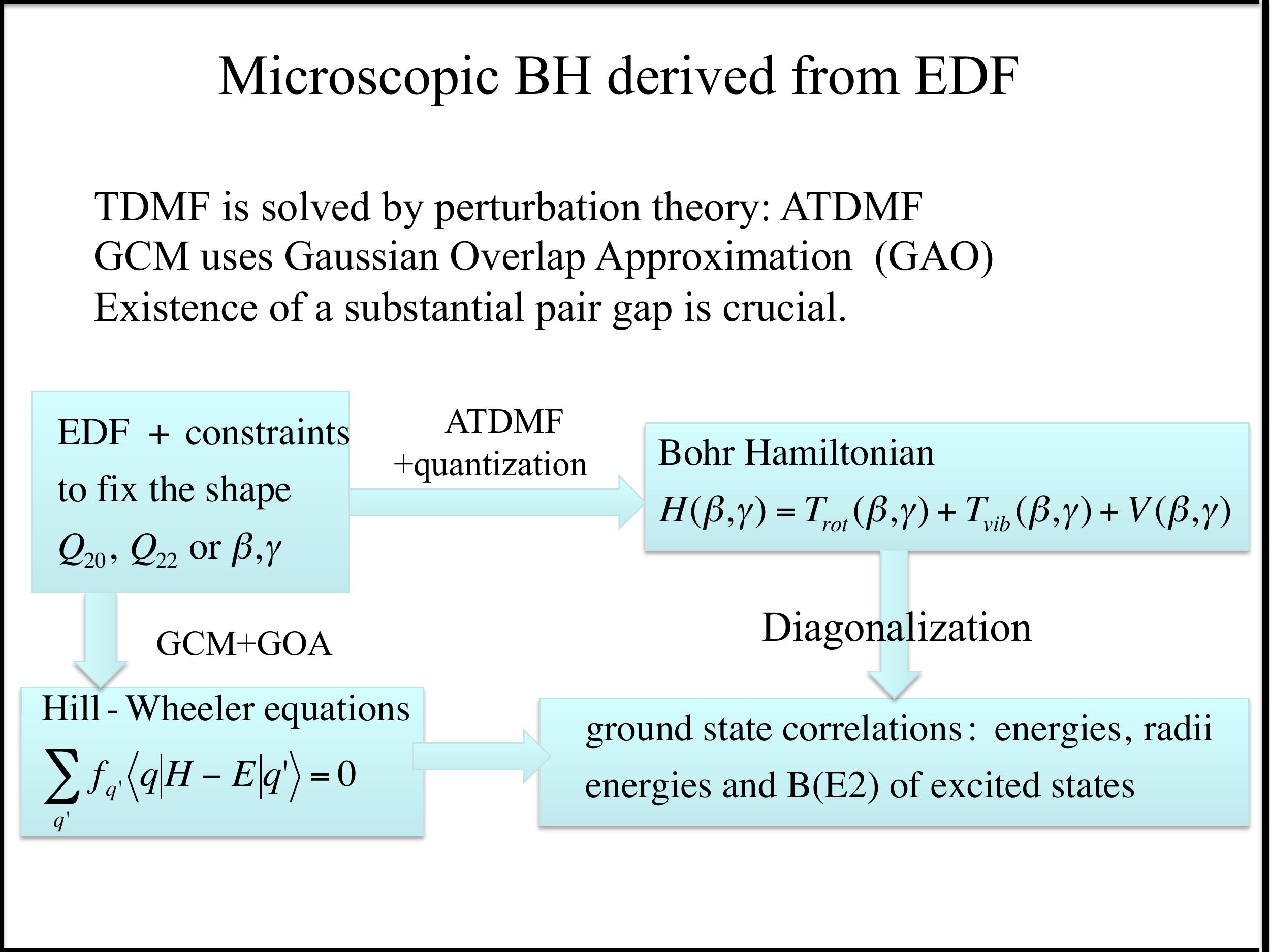}\includegraphics[width=0.5\linewidth]{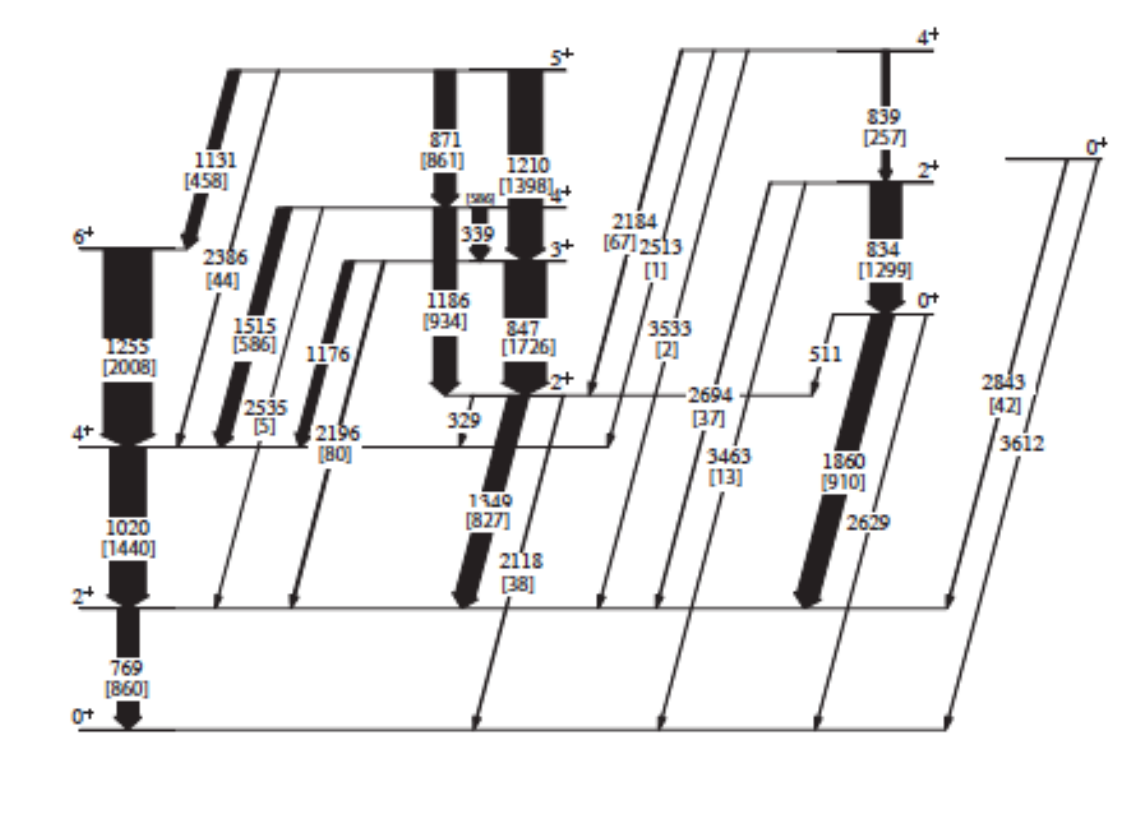}}
\caption{\label{fig:5DBH} Left: Scheme of microscopic derivations of the Bohr Hamiltonian.
 Right: The spectrum of $^{102}$Pd. 
calculated by means of the 5DBH-GI \cite{Delaroche10}. 
The number in parenthesis under the transition energies (keV) are the $B(E2)$ values ($e^2$ fm$^4$) for the transitions. From Ref. \cite{Frauendorf15}.}
\end{figure}
\subsection{Microscopic Bohr Hamiltonian}
The GCM is a useful way to classify the collective quadrupole excitations of even-even nuclei. To predict their properties, the BH is derived 
from the  nuclear mean field. The derivation and subsequent calculation of the collective excitations has been carried out for all popular mean field approaches. 
 The derivation of the BH from the various mean field approaches has been detailed reviewed  in Ref. \cite{Frauendorf15}.
Fig. \ref{fig:5DBH} left summarizes the work in  a schematic way.  The potential $V(\beta,\gamma)$  is  the energy of constrained mean-field solutions generated from
the different energy density functionals (EDF). The kinetic energy  $T$ is obtained in two different ways. 
The  adiabatic time dependent  mean field approach (ATDMF) calculates the increase of the energy when the shape is slowly changed via time-dependent constraints, where
only the leading term of time-dependent perturbation theory is kept. The use of perturbation theory indicates the adiabatic nature of the approach. The second
approach starts from the Hill-Wheeler equation  for the weight function  of the Generator Coordinate Method (GCM), which describes the collective
wave function as a superposition of the different constraint mean-field solutions. Approximating  the overlap between the non-orthogonal  mean-field solutions
by a Gaussian (GOA) allows one to cast the Hill-Wheeler equation into the differential form of the BH. It needs to be underlined that the restriction to the 
zero quasiparticle solutions in the GCM and the use of  of the  GAO implicitly  introduce an adiabatic approximation. The BH  and its 
eigenfunctions are obtained  numerically on a deformation grid. 

 Large scale calculations of the lowest collective excitations have been carried out for various commonly used mean-field approaches
  (see review \cite{Frauendorf15}). 
  In a bench mark study, Ref. \cite{Delaroche10} carried out calculations for all even-even
with  $10\leq Z\leq 100$ and $20 \leq A\leq 200$. The results for the energies and E2 and E0
matrix elements for the yrast levels with $ I \leq 6$, the lowest excited $0^+$ states, and the
two next yrare $2^+$ states are accessible in the form of a table as supplemental material
to the publication. Fig. \ref{fig:5DBH} right shows $^{102}$Pd as an example,
which illustrates the typical accuracy of the predictions. 
A thorough statistical analysis of the merits of performance has
been carried out. The authors state: "Many of the properties depend strongly on the
intrinsic deformation and we find that the theory is especially reliable for strongly
deformed nuclei. The distribution of values of the collective structure indicator
$R42 = E(4^+_1 )/E(2^+_1 )$ has a very sharp peak at the value 10/3, in agreement with the
existing data. On average, the predicted excitation energy and transition strength
of the first $2^+$ excitation are 12\% and 22\% higher than experiment, respectively,
with variances of the order of 40-50\%. The theory gives a good qualitative account
of the range of variation of the excitation energy of the first excited $0^+$ state, but
the predicted energies are systematically 50\% high. The calculated yrare $2^+$ states
show a clear separation between $\gamma$ and $\beta$ excitations, and the energies of the $2^+,~ \gamma$
vibrations accord well with experiment. The character of the $0^+_2$ state is interpreted
as shape coexistence or $\beta$-vibrational excitations on the basis of relative quadrupole
transition strengths. Bands are predicted with the properties of $\beta$ vibrations for
many nuclei having R42 values corresponding to axial rotors, but the shape coexistence
phenomenon is more prevalent."

 In addition they observe that the $0^+_2$ states are generally Òtoo vibrationalÓ. The discrepancy
 is particularly precarious for well deformed nuclei. For example, the  $0^+_2$ states in $^{166}$Er and
 $^{178}$Hf do not show the enhanced E2 transition probability expected for a $\beta$ vibration, which is  predicted by the BH.
 Instead some collective enhancement for transitions from higher $0^+$ states is observed, which indicates the fragmentation 
 or absence of the collective $\beta$ vibration \cite{Garrett97,Lesher07}.

\section{INTERACTING BOSON MODEL}\label{sec:IBA}
The collective quadrupole mode is described by means of 
the creation operators for nucleon pairs with spins 0 and 2, which are called s- and d- bosons and denoted by $s^{\dagger}$ and $d^{\dagger}$,
respectively, form the closed Lie algebra of the SU(6) group. 
In the simplest version, the Hamiltonian  contains  two IBM parameters and the energy scale, such that:
\begin{equation}
\label{eqn:HIBM}
H_{IBMA}(\zeta , \chi)= g\Big((1-\zeta) \hat{n}_d-\frac{\zeta}{4N_B} \hat{Q}^{\chi}\cdot\hat{Q}^{\chi}\Big),
\end{equation}
where $\hat{n}_d=d^{\dagger}\cdot\tilde{d}$, and $\hat{Q}^{\chi}_\mu=[s^{\dagger}\tilde{d}+d^{\dagger}s]^{(2)}_\mu+\chi[d^{\dagger}\tilde{d}]^{(2)}_\mu$.
The Hamiltonian is diagonalized within the space of  fixed number of bosons, $N_B=n_s+n_d$, which is taken to be
 half the number of valence nucleons. 
 The matrix elements of the charge quadrupole moments are taken to be proportional to $\hat{Q}^{\chi}$, with an effective boson 
 charge fixing the scale. As in the case of the GCM, the two parameters $\zeta$ and $\chi$ determine the character of the collective states. 
The two-parameter fits usually well account for the relative energies of the   $2^+_1$, $4^+_1$, $6^+_1$, $2^+_2$, $3^+_1$, $4^+_2$, $0^+_2$, $4^+_3$ states
and the relative $B(E2)$ for the transitions between them. The quality of the fits is comparable with the two-parameter version of the GCM.

 \begin{figure}[t]
\centerline{\includegraphics[width=0.5\linewidth]{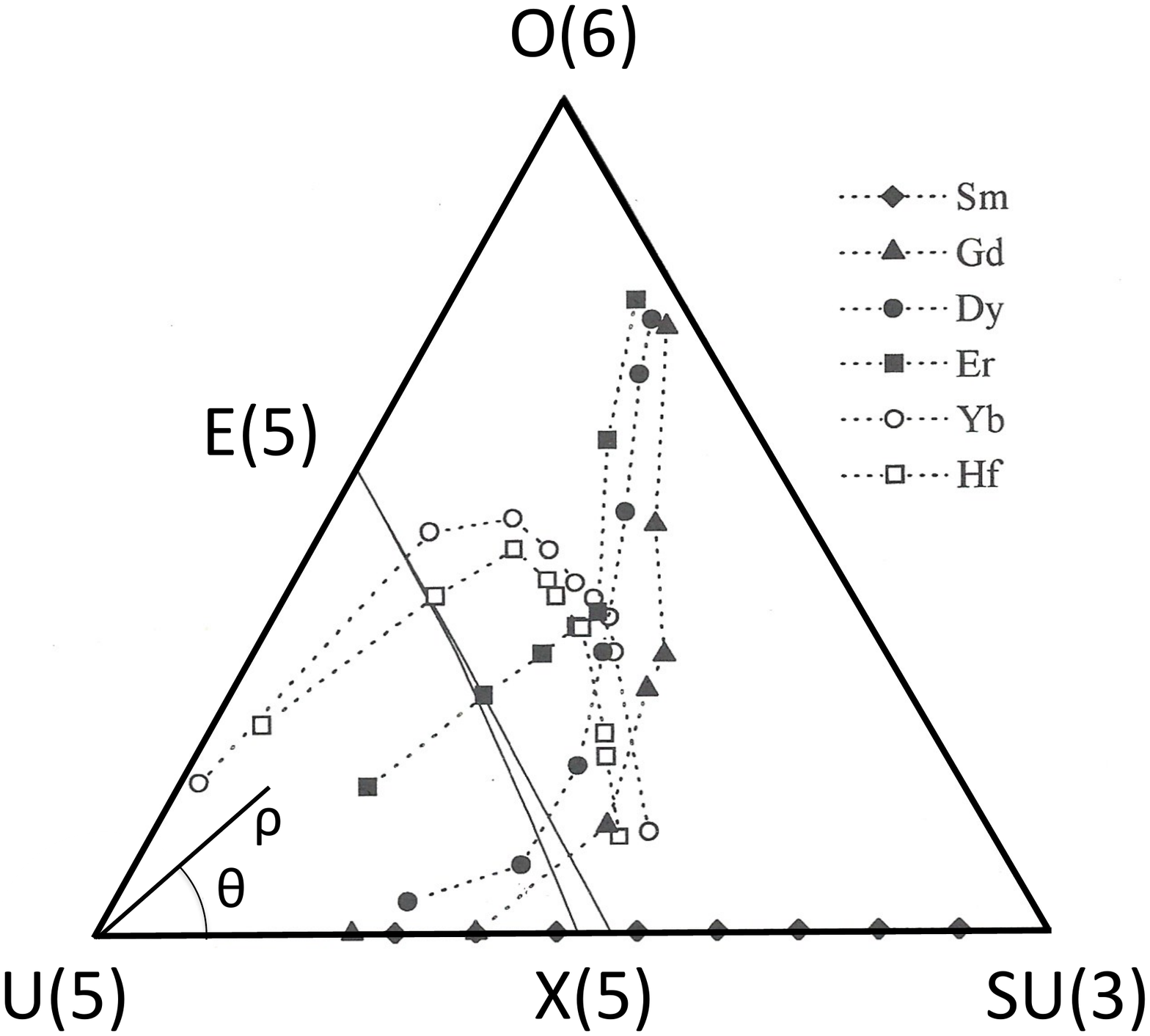}\includegraphics[width=0.5\linewidth]{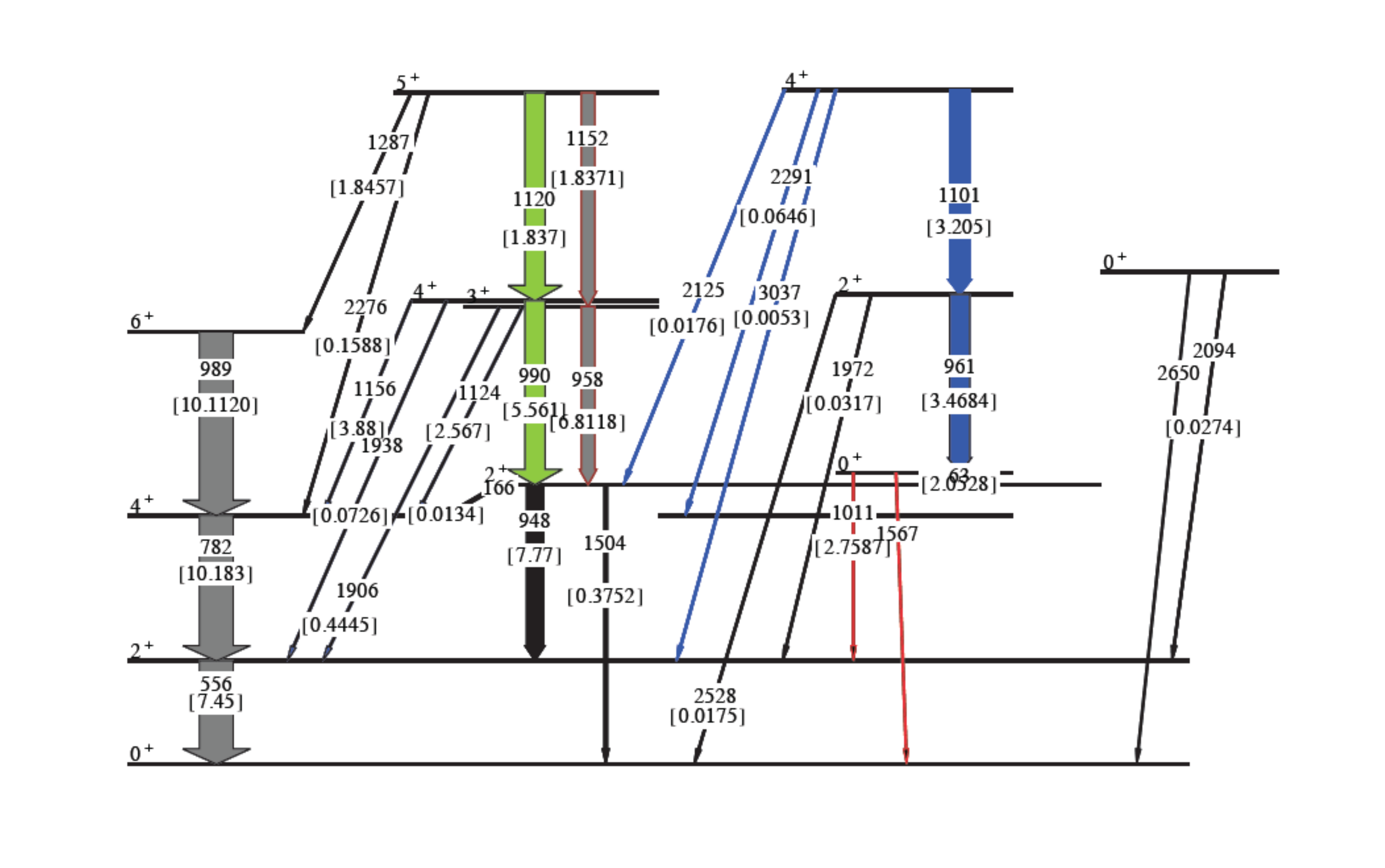}}
\caption{\label{fig:102PdIBM} Left:The IBM parameters of selected isotope chains arranged in the Symmetry Triangle. The symmetry limits
of the subgroups are indicated at the three corners. The  "transition point symmetry" X(5) marks 
the transition from the vibrational regime (limit U(5))  to axial deformation (limit SU(3)  
and E(5) marks the transition to the $\gamma$ - soft deformed regime (limit O(6)). Right:  IBM-1 fit to the spectrum of $^{102}$Pd.   
The number in parenthesis under the transition energies (keV) are the $B(E2)$ values ($e^2$ fm$^4$) for the transitions.
The IBM parameters are $\zeta=0.61,~\chi=0.58, N_B=5$.  From Ref. \cite{Frauendorf15}
}
\end{figure}

It has become custom to classify the collective states on the "symmetry triangle", which is a polar plot generated from the two parameters $\zeta$ and $\chi$.
The three corners of the triangle are parameter combinations that generate additional symmetries with respect to the subgroups U(5), O(6), SU(3), 
 which correspond to  the harmonic vibrator, the $\gamma$ -independent, axial rotor limits of the GCM approach. It is an attractive feature of the IBM that
 simple algebraic expressions describe the energies and reduced transition probabilities of the three symmetry limits. Examples are displayed in Fig. \ref{fig:102PdIBM}.
  Left shows the classification of some isotope chains. Right demonstrates the quality of IBM for $^{102}$Pd. Comparing with Fig. \ref{fig:102PdACM} 
  demonstrates that both the GCM and the IBM reproduce the low-spin part of the spectrum equally well.
 
 \begin{figure}[h]
\centerline{\includegraphics[width=0.5\linewidth]{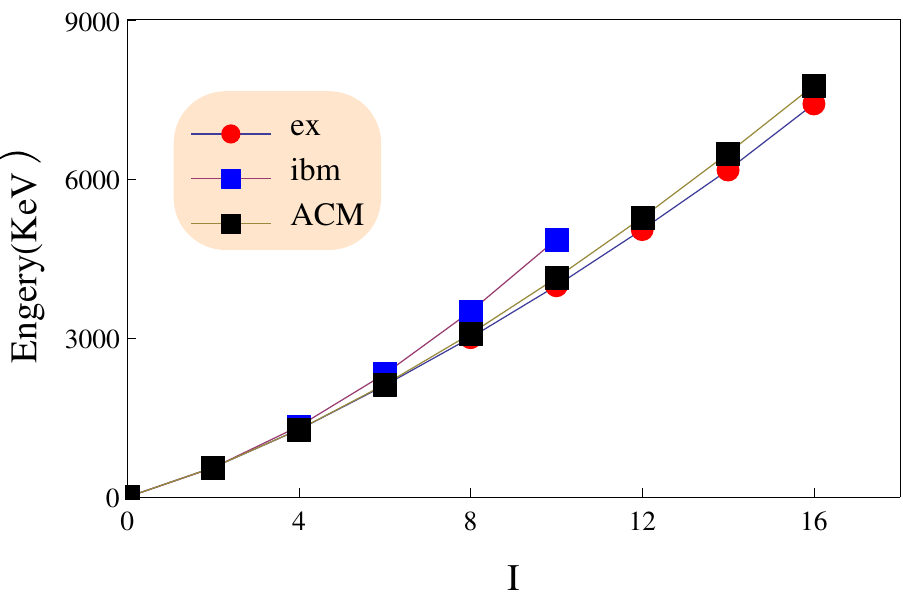}\includegraphics[width=0.5\linewidth]{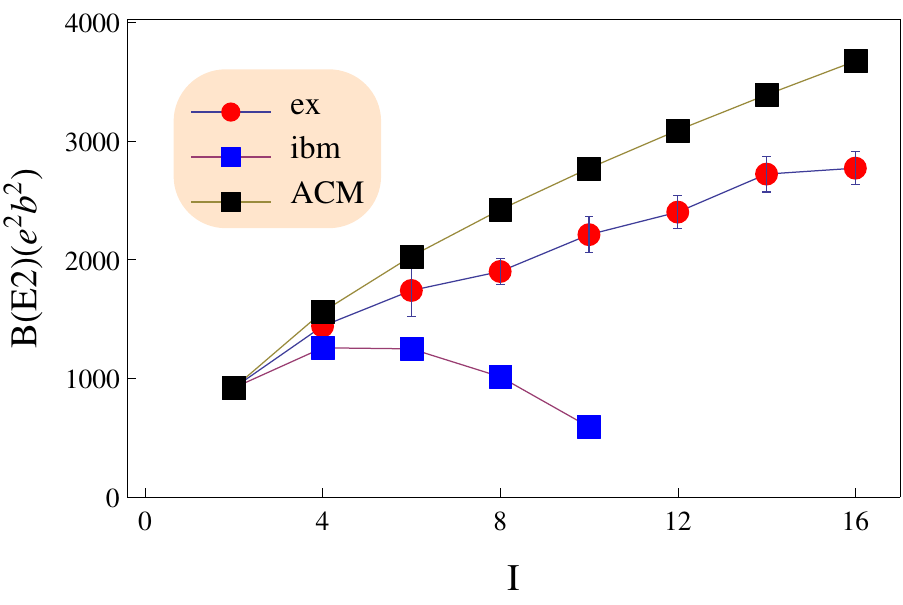}}
\caption{\label{fig:102PdEBE2} Energies (left) and $B(E2, I\rightarrow I-2)$ values (right) of $^{102}$Pd.  The GCM calculation, denoted by ACM, is the same as the one shown in
Fig. \ref{fig:102PdACM} and the IBM calculation is the same as in Fig. \ref{fig:102PdIBM}. Data from Ref. \cite{102PdTidalPRL}.} 
\end{figure}   
  
The IBM has been conceived as an approximation to the Shell Model. The configuration
space of the valence nucleons between two closed shells is truncated  to the subspace of pairs coupled to spin
zero and two, which are then mapped to the space of the $s$ and $d$ bosons. The number $N_B$  of such pairs is  taken
one half of the number of valence particles below the middle of the open shell or one half of the valence holes above the middle.
The finiteness of the boson number is  the major difference between  the IBM and the GCM. The consequences of the boson number
limit are not well studied because the IBM is usually  applied to  low-spin states that correspond to bosons numbers far below the limit.  
The recent measurement of the life times of the yrast states in $^{102}$Pd in Ref. \cite{102PdTidalPRL} provides such test. 
As illustrated by Fig. \ref{fig:102PdEBE2},
the yrast levels form a regular sequence of collective excitations with increasing $B(E2, I\rightarrow I-2)$ values up to $I=14$. 
The data are compared with the GCM and IBM calculations that give the spectra shown in Figs. \ref{fig:102PdACM} and  \ref{fig:102PdIBM}. According to the IBM counting rule,
$^{102}_{46}$Pd$_{56}$ has a boson number of $N_B=5$: four valence  proton holes and six valence neutrons with respect to $Z=N=50$. The  boson number 
limits the regular yrast sequence at $I=10$, which is the maximum that can be generated by five  $d$ bosons. The $B(E2)$ values decrease toward
the limit where they are zero. These consequences of the finite boson number are in clear contradiction with experiment. The GCM calculation, which does not assume a boson limit
reproduces the experiment quite well. On the other hand,
 Figs. \ref{fig:102PdACM} and  \ref{fig:102PdIBM} demonstrate that both the GCM and the IBM describe  the spectrum comparably well.
The study of E2 transition matrix elements in $^{168}$Er provides evidence that the IBM systematically underestimates their collectivity  for the highest spins
as a consequence of the boson number cut-off \cite{Er168ME2}.  Although these are only two examples, they may indicate a general problem of the IBM.
Testing the consequences of the finite boson number assumption in a systematic way would be an important test of the corner stone of the IBM. 

IBM assumes that the $s$ and $d$ bosons 
 represent valence nucleon pairs in spherical orbitals that are coupled to spin 0 or 2.  A microscopic derivation  of the IBM parameters starting from this  concept has not
 been succeeded  for nuclei located far in the open shell. 
An alternative approach has provided 
encouraging results. The IBM parameters are determined by adjusting an IBM potential energy surface  $E_{IBM}(\beta_{B},\gamma_B)$ generated from
the coherent-state representation of the IBM Hamiltonian (\ref{eqn:HIBM}) 
to the potential energy surface  $E_{mf}(\beta_F,\gamma_F)$ calculated by means of constraint mean field theory. For references and an extended presentation see Ref. \cite{Frauendorf15}.

\section{ROTATING MEAN FIELD}\label{sec:RMF}

 Although the excitation energy is high for large spin,  the levels density remains low near the yrast line.
There,  uniform rotation prevails, which  is the yrast solutions  of the BH. Uniform rotation can be studied  in a non-adiabatic way by means 
of the rotating mean-field approximation. The method is used for all versions of modern mean-field approaches.  
In essence, one finds quasiparticle configurations that are generated by the quasiparticle Routhian 
\begin{equation}\label{eq:h'sp} 
 h'(\beta,\gamma,\omega)= h(\beta,\gamma) -\vec \omega \cdot  \vec j, ~~~h_{qp}=h(\beta,\gamma)+\Delta(P^\dagger+P)-\lambda \hat N,
 \end{equation}
 The cranking term $\vec \omega \cdot \vec j$ transforms to the frame that rotates with the angular velocity $\vec \omega$. The rotational axis is usually taken as one of the principal axes  
 the deformed single particle Hamiltonian $h(\beta,\gamma)$. However for a class of solutions the rotational axis titled with repeat to the principal axes (Tilted Axes Crankinng).    
 Each quasiparticle configuration is associated with  a rotational band.
   The details of deformed single particle Hamiltonian depend on the mean-field approach of choice. Simple and popular 
 are the Woods-Saxon or the Nilsson potentials, which are combined with the shell correction  (or micro-macro)  method  
 to calculate the energy in the rotating frame $E'(\beta,\gamma,\vec \omega)$ and the expectation value of the angular momentum $J( \beta,\gamma,\vec \omega)$.
 The deformation parameters $\beta, \gamma$ and the orientation of the rotational axis $ \hat {\vec\omega}$ are found be minimizing the energy 
 $E(\beta, \gamma,\ \hat {\vec\omega})=E'+\omega J$ under the constraint of fixed angular momentum $J$.  The pair-field $\Delta$ is  determined
 by various prescriptions and $\lambda$ fixes the particle number. Semiclassically,  the intraband transition matrix elements are obtained by calculating the expectation
 values of the transition operators, as e.g. the quadrupole moments.  
 In my talk a I can only address few aspects of this wide field. Recent reviews, which expose the details,  have been given by  Frauendorf \cite{RMP} and Satu\l a and Wyss \cite{WS05}. 
  \begin{figure}[t]
\centerline{\includegraphics[width=0.5\linewidth]{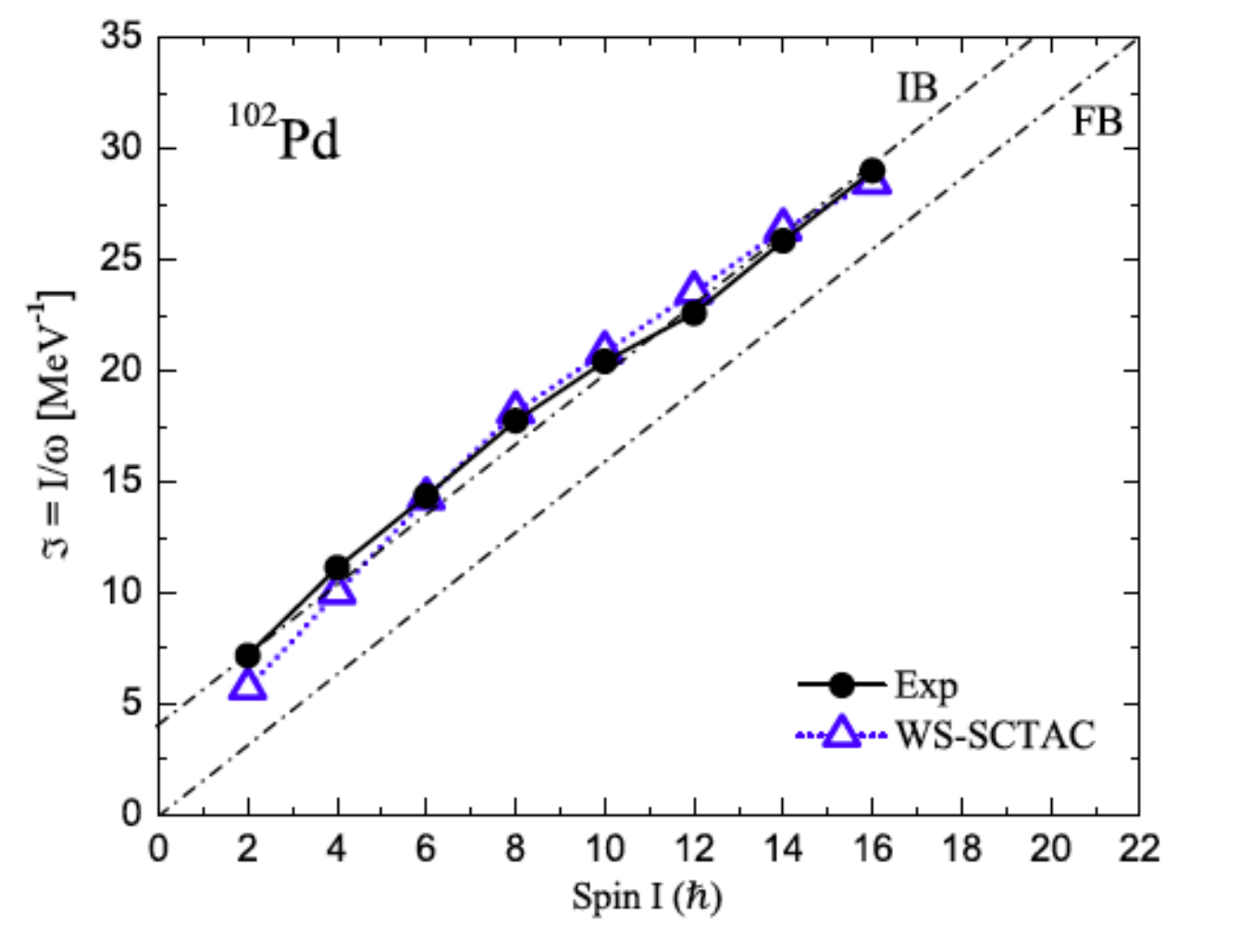}\includegraphics[width=0.5\linewidth]{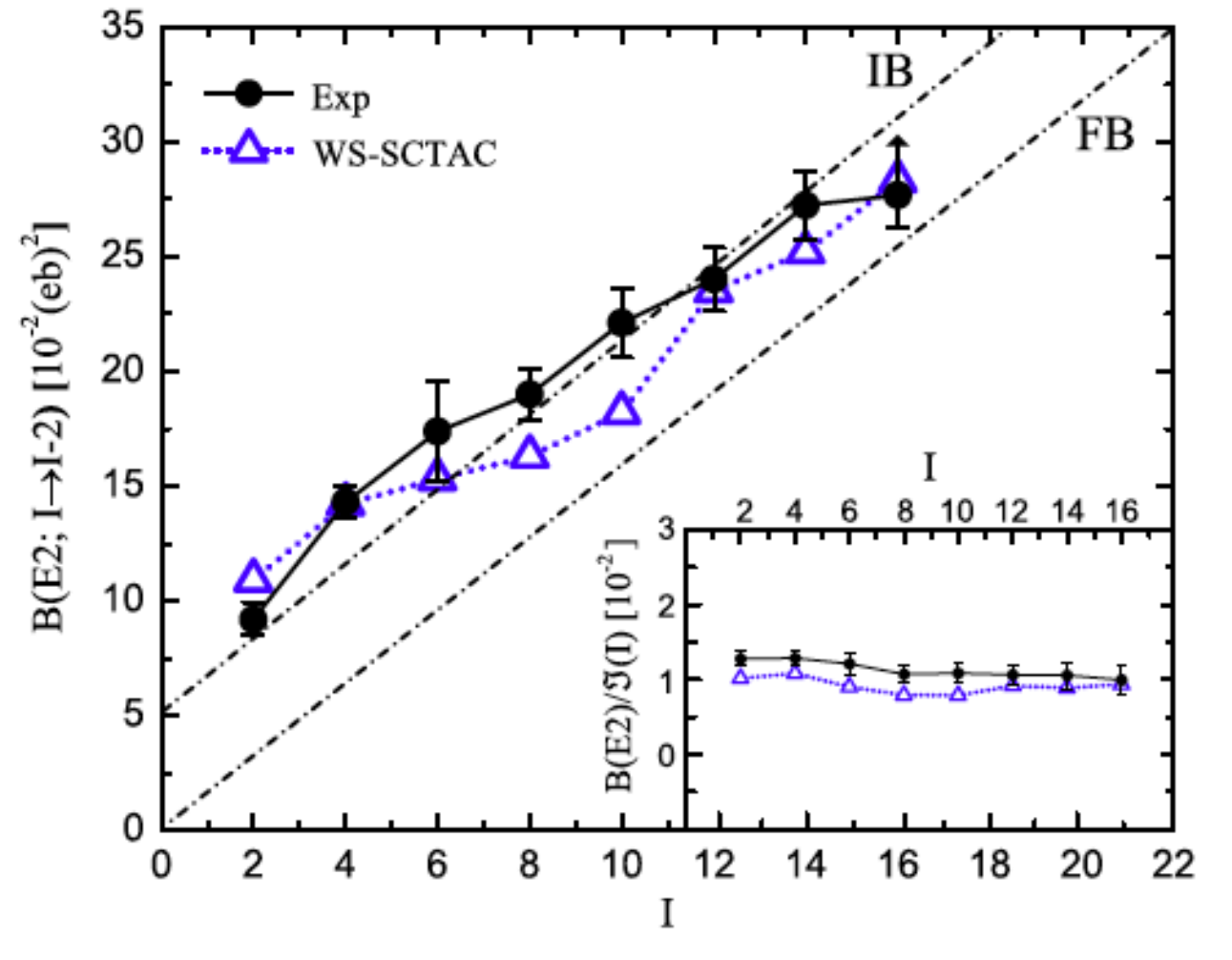}}
\caption{\label{fig:102PdPRL} 
Left part: Experimental moments of inertia of the ground band 
of $^{102}$Pd, where \mbox{$\omega=(E(I)-E(I-2))/2$} and $J=I$.  Right part:  Experimental $B(E2, I\rightarrow I-2)$ values of $^{102}$Pd.
The blue lines WS-SCTAC show the calculations by means of the cranking model \cite{FGS11}. From Ref. \cite{Frauendorf15}.  } 
\end{figure}

 The rotating mean-field (RMF) opened new perspectives because it is non-perturbative and completely microscopic. It allows one to address the question of the 
 appearance of the  rotational modes away from closed shells and their disappearance with excitation energy above the yrast line.

 \subsection{Tidal Waves} 
Frauendorf, Gu and Sun  introduced the Tidal Wave concept in Ref. \cite{FGS11} (and Ref. \cite{FGS10} with  complimentary material). 
 Consider the phenomenological  BH    (\ref{eqn-HGCM2}) with mass parameters  $B_{\beta\beta}=B_{\gamma\gamma}=B_i=\sqrt{5}/2D$.  
 Uniform rotation about the  axis with the maximal moment of inertia has the lowest energy for a given angular momentum, i. e. it corresponds to the yrast states when quantized. 
In the co-rotating frame, the deformation parameters $\beta$ and $\gamma$ do not depend on time. 
 Their values are given by minimizing the energy 
 \begin{equation}\label{Ebeta}
E(\beta,\gamma)=\frac{J^2}{2{\cal J}(\beta,\gamma)}+V(\beta,\gamma),~~
{\cal J}=4B\beta^2\sin^2\gamma.
\end{equation}
In the case of a harmonic vibrator  $V=\frac{C}{2}\beta^2$. Minimizing the energy one finds
\begin{eqnarray}
\gamma_e=\frac{\pi}{2},~~~ \beta^2_e=\frac{J}{2\sqrt{BC}},~~~ {\cal J}=4B\beta^2_e=\frac{2J}{\sqrt{BC}},~~~
\omega=\frac{J}{{\cal J}}=\frac{1}{2}\sqrt{\frac{C}{B}},~~~E=\omega J=\Omega\frac{J}{2}=C\beta^2_e.
\end{eqnarray}
The wave travels with an angular velocity $\omega$ being one half of the oscillator frequency
$\Omega$. The angular momentum is generated by increasing the deformation $\beta^2$, while the angular velocity stays constant.
These are the yrast states  of the vibrator multiplets described in a semiclassical way. 
This mode has been called  "Tidal Wave" (TW), because it has wave character: the energy and angular momentum increase with the wave amplitude while
the frequency stays constant. Using a potential like  Eq. (\ref{eqn-VGCM2}) one can easily incorporate 
anharmonicities and cover  the transition to stable rotation.

 \begin{figure}[h]
\centerline{\includegraphics[width=0.5\linewidth]{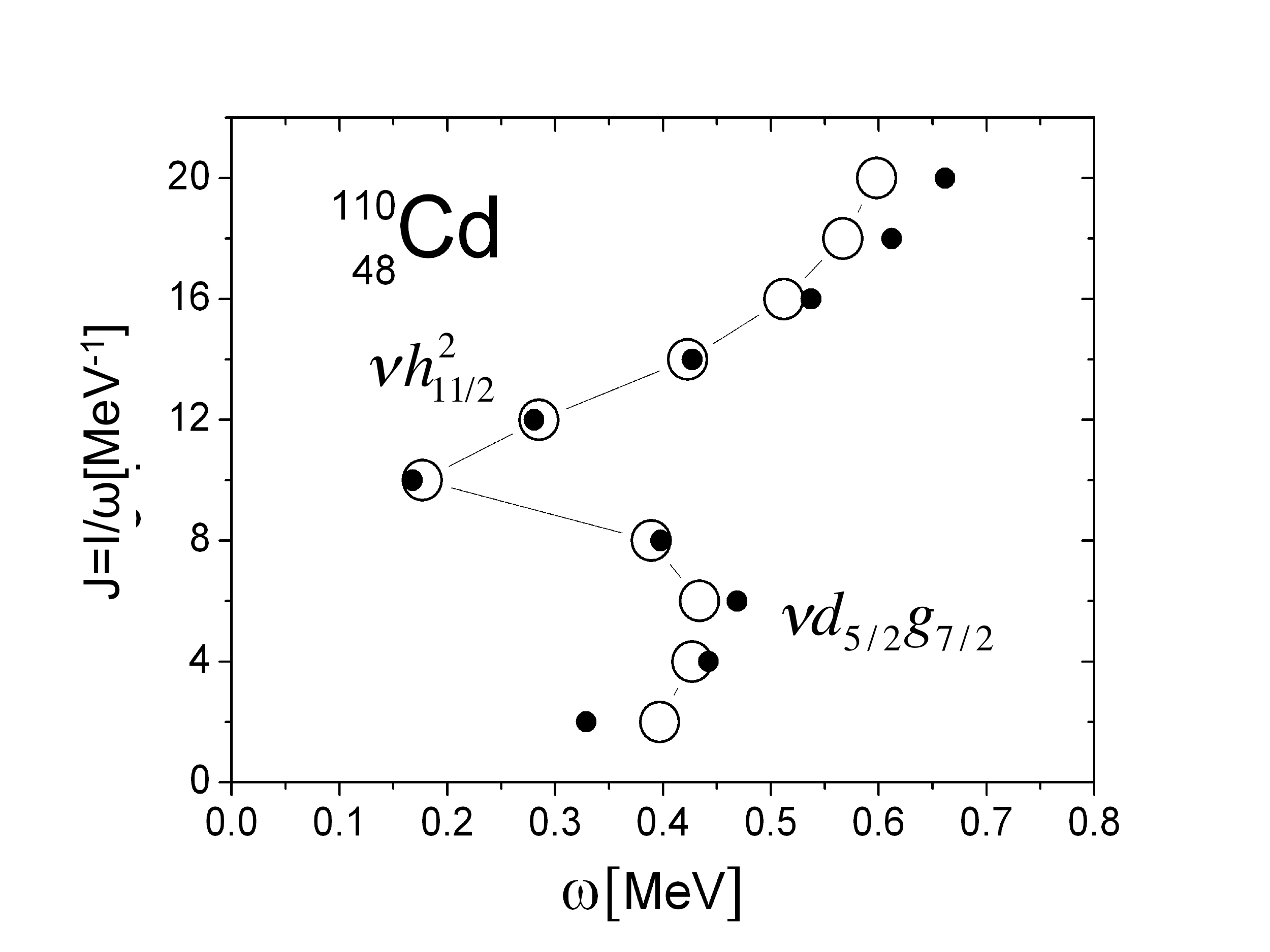}\includegraphics[width=0.5\linewidth]{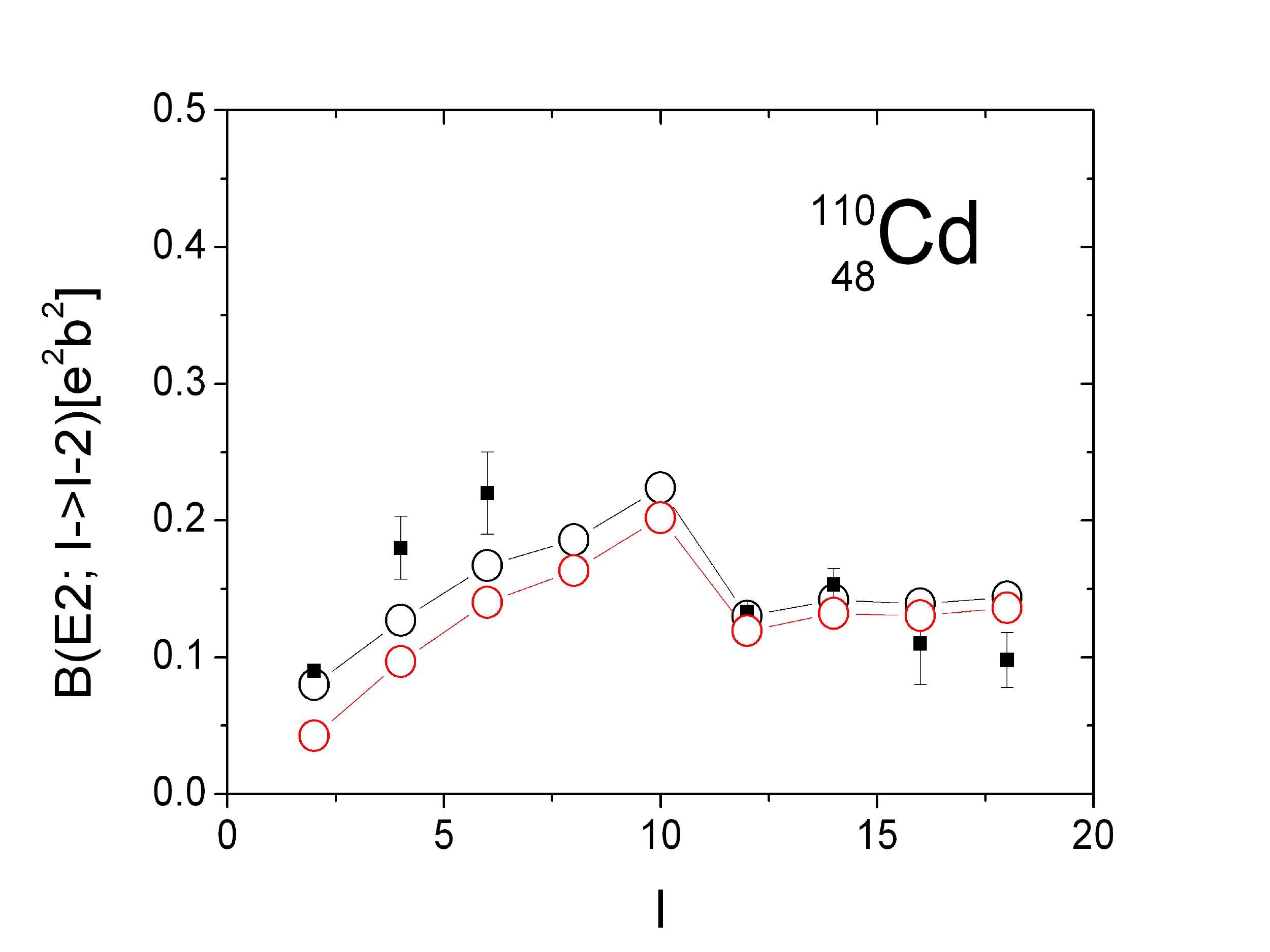}}
\caption{\label{fig:110CdTAC} 
Left part: Experimental moments of inertia of the ground band and 0$^+_2$ band (black dots)
of $^{110}$Cd, where \mbox{$\omega=(E(I)-E(I-2))/2$} and $J=I$ compared with the  calculations by means of the Cranking model \cite{FGS11} (black circles).
 Right part:  Experimental $B(E2, I\rightarrow I-2)$ values of $^{110}$Cd compared with the  calculations (black circles without red circle with quantal correction, see Ref.
 \cite{FGS11}). From Ref. \cite{Frauendorf15}.  } 
\end{figure}    

The yrast states of $^{102}$Pd shown in Fig. \ref{fig:102PdEBE2} are a beautiful example of a slightly anharmonic TW. 
Ayangeakaa {\it et al.}\cite{102PdTidalPRL}  interpreted it as a condensate of $d$ bosons. 
Accordingly, up to seven bosons are observed, which align their angular momenta. 
If the bosons were free, the function 
${\cal J}(I)$ would be a straight line out of the coordinate origin (FB in Fig. \ref{fig:102PdPRL}). The displacement by $\Theta_0$  
was attributed  to an interaction between the bosons (IB in Fig. \ref{fig:102PdPRL}) that is quadratic in the boson number \cite{MAFGC14}.

The fact that the TW is static in the co-rotating frame of reference allows one to microscopically calculate its properties by means of the RMF approaches. 
Frauendorf, Gu, and Sun \cite{FGS10,FGS11} calculated the energies of the yrast states and the $B(E2)$ of the intra band transitions up 
to spin $I=16$ for the nuclides with $Z=44-48,~N=65-66$.  
Figs. \ref{fig:102PdPRL} and \ref{fig:110CdTAC} exemplify the accuracy of the parameter-free calculations. In particular the change of the yrast states from the 
purely collective tidal wave (g band) to the  configuration  with two rotational aligned h$_{11/2}$quasiparticles  (s band) is reproduced in detail. 
In the case of $^{102}$Pd (Fig. \ref{fig:102PdPRL}) the collective g  band can be followed up to $I=14$, where it is at higher energy than the s  band. 
This is a consequence of almost no mixing of the two configurations.
In the case of $^{110}$Cd (Fig. \ref{fig:110CdTAC})  the collective g band is crossed by the s band earlier and the two bands interact stronger. The
two aligned h$_{11/2}$ quasiparticles  in the s band reduce the deformation but stabilizes it such that the sequence becomes more rotational. 
The method applies to odd-A and odd-odd  nuclei without any further sophistication. 

\subsection{Coherence Of Rotational Motion}
 The nucleus increases its angular momentum in two different ways. One is coherent rotation of the nucleus, which results in regular rotational bands.
 The other is exciting  quasiparticles that align they individual angular momentum, which appear in an irregular way. Fig. \ref{fig:110CdTAC} shows 
 an example for the competition of the two modes.   The RMF accounts for both on equal footing. It is important to realize that the quasiparticle Routhian (\ref{eq:h'sp})
 derives from an effective two-body Routhian that is invariant with respect to rotation about the $\vec \omega$ axis. Rotational invariance implies that there is a family 
 RMF solutions related by rotation about $\vec \omega$ by the angle $\psi$.   All of them have the same energy.  In the space fixed coordinate system these mean field states 
 rotate  uniformly about the $\vec \omega$ axis. In the spirit of  semi classical quantum mechanics this motion is quantized, i. e.  wave functions are 
 associated with the angle that specifies the orientation 
  of the degenerate mean-field solutions. They are the microscopic realization of  the collective wave functions of a rotor.  
   The orientation angle $\psi$ exists only if the mean-field solution breaks rotational symmetry with respect to the   $\vec \omega$ axis. 
  One calls this "spontaneous symmetry breaking" because the original two-body Routhian preserves the symmetry. 
  
  To quantify the degree of symmetry breaking it is instructive to introduce the notation of a "coherence length",
  which is used in other fields of many-body physics. It is the minimal length that a collective wave function can resolve. In the case of superconductivity the coherence length
  $\xi=\pi\De/p_f$ is the size of a Cooper pair. The wave function of the pair condensate cannot more rapidly change than $\xi$.  When the pair condensate flows through a wire,
  its wave function acquires the phase $ipx/\hbar$. The phase cannot change more rapidly than $\xi$, that is $p<p_{max}=\hbar/\xi$. When the  current through the wire
  is increased, superconductivity breaks down at the critical current density \mbox{ $j_{max}=e\rho p_{max}/2m=10^3-10^4 A/m^2$}. 
  
   In analogy, there exists a coherence angle $\Delta \psi$ that limits the  resolution 
  of the  wave functions of collective rotation, which
   are called "coherent" because the nucleons are correlated such that a common orderly motion results. The coherence angle can be determined from the overlap 
  between the different mean-field solutions $\vert \psi\rangle$  that specify the angle $\psi$. The overlap is usually well approximated by a Gaussian, 
   \mbox{$\vert \langle \psi\vert \psi'\rangle\vert^2\approx\exp \left[-(\psi-\psi')^2/\Delta \beta^2\right]$}, which allows one to calculate  $\Delta \psi$.
  The correlation angle sets the limit how much phase $\psi J/\hbar $   the rotational wave can function acquire and  such it restricts the number of states in the rotational band to $\Delta I\sim2/\Delta \psi$.
  ( The 2 is chosen such that the number of band members correspond to the observed ones)
  Ref. \cite{RMP} provided an extended of discussion of the coherence of rotation, which  formulated  additional criteria for the appearance of regular rotational bands.   
   
  \begin{figure}[h]
\begin{minipage}[t]{0.5\textwidth}
\hspace*{1cm}
\begin{tabular}{|crrr|}
deformation&super &normal &weak\\
\hline
${\cal J}^{(2)}$&97&56&14\\
$\Delta J$&28&14&6\\
$\Delta \psi $&4$^o$&8$^o$&20$^o$\\
$Q_t$&5.2&2.6&0.7\\
$\mu_t$&0&0&3.5\\
\hline
$Z$&64&72&82\\
$N$&88&104&117\\
$\eps$&0.6&0.3&0.1\\
$\De_p$&0&0.75&0\\
$\De_n$&0&0.70&0.75\\
\end{tabular}
\end{minipage}
\begin{minipage}[t]{0.5\textwidth}
\vspace*{-2.3cm}
\includegraphics[width=\linewidth]{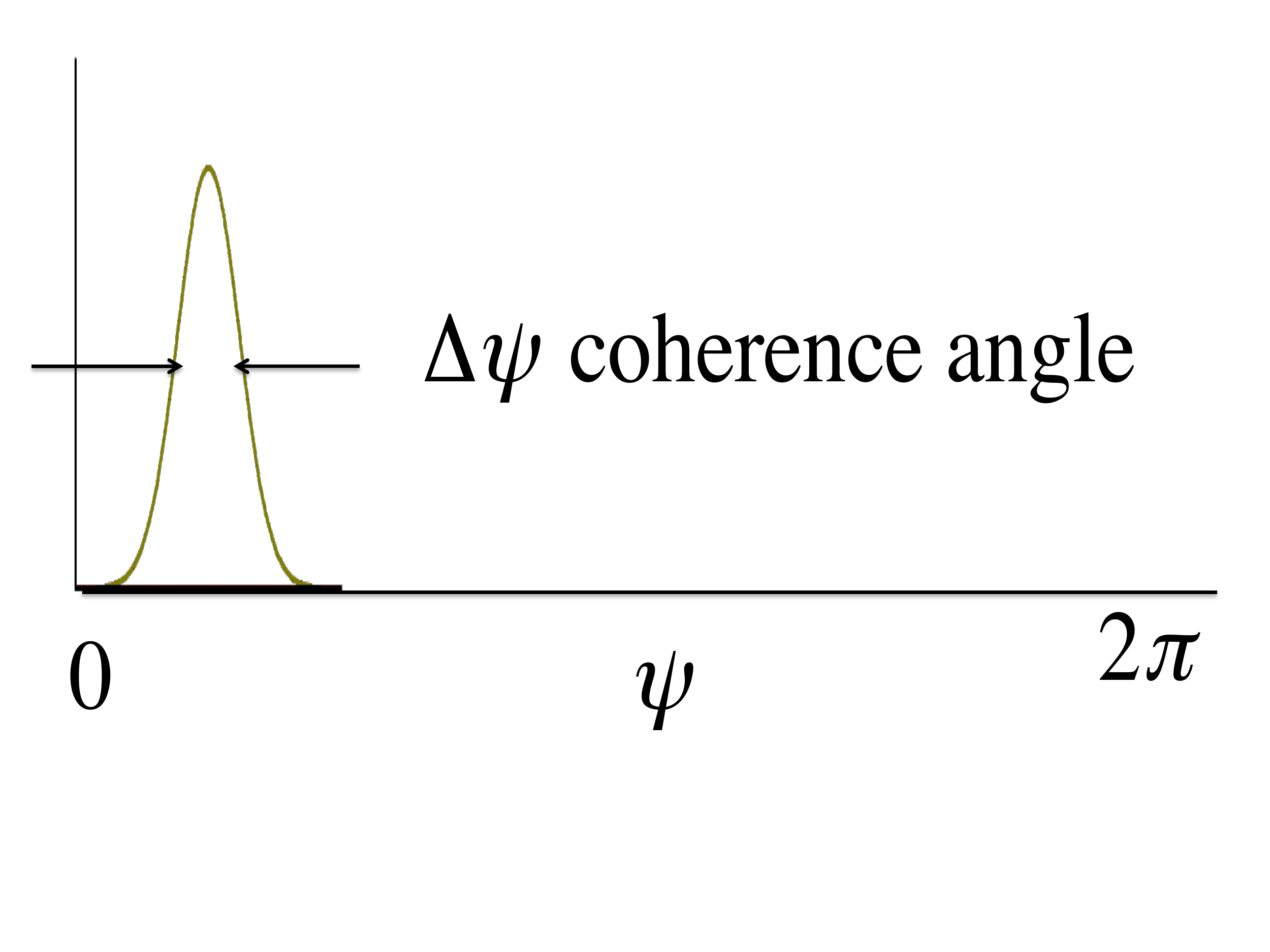}
\end{minipage}
\caption{\label{fig:CohAng}
Left: Character of the  different types of nuclear  rotational bands.
The moments of inertia ${\cal J}^{(2)}$ are in  $\hbar^2MeV^{-1}$,
the transition quadrupole moments $Q_t$ in $eb$, the transition magnetic moments $\mu_t$ in $\mu_N$ and
the pair gaps $\De$ in $MeV$. The parameter $\eps$ measures the deformation ($\eps\approx0.95\beta$ for $\beta < 0.2$).
 Right: Numerical calculation of the overlap for the case of weak deformation. From \cite{RMP}.}
\end{figure}

 The table in Fig.   \ref{fig:CohAng} provides  three examples of different types of rotational bands found in nuclei. The super deformed   deformed nuclei are found at high spin.  
 They are characterized by long regular bands. The normal deformation case illustrates how the finite length of the bands materializes in well deformed nuclei
 (e. g. the  rare-earth region). 
  The regular level sequence of the state band stops 
  around $I=12$, where it is crossed by another band  containing a pair of $i_{13/2}$ neutrons that aligned their spins with the rotational axis. 
  The alignment causes an sudden decrease of the transition energy, called back bending. 
  Crossings with bands containing an increasing number of aligned quasiparticles recur at intervals of the order of $\De I$. 
 
  The weak deformation case is an example for magnetic rotation in the near-spherical Pb isotopes.   Magnetic rotation is carried by few high-j nucleons 
  on orbits that are arranged in a highly anisotropic way, which results in the largest coherence angle of the three examples.  As seen in Fig. \ref{fig:CohAng}, it is still
  a small fraction of $2\pi$,  enough to support phase changes of $\sim    2\pi/6$.  As indicated by the small value 
  of $Q_t$ and the large value of $\mu_t$ in the table of Fig. \ref{fig:CohAng},
  the rotational  band appears as a regular sequence of strong M1 transitions while the E2 transitions are suppressed.
  
  Band termination is a second phenomenon how the finite length of  bands materializes in transitional nuclei. Along the band
  the  valence nucleons gradually align their spins with the rotational axis. 
  The band terminates when all spins are aligned in accordance  with the Pauli Principle. The terminating configuration is spatially symmetric with
  respect to the angular momentum, e. g not E2 radiation is possible.  The $B(E2)$ values  fall off  while approaching  termination along the band.
  Termination competes with band crossings with varying prevalence.

   Magnetic rotation and band termination are a phenomena beyond the classical UM, which appear naturally as consequence of spontaneous symmetry breaking
   of the rotating mean field. Ref. \cite{RMP} discusses these aspects in detail together with the consequences of spontaneous breaking of discrete symmetries.

  \begin{figure}[t]
\centerline{\includegraphics[width=0.48\linewidth]{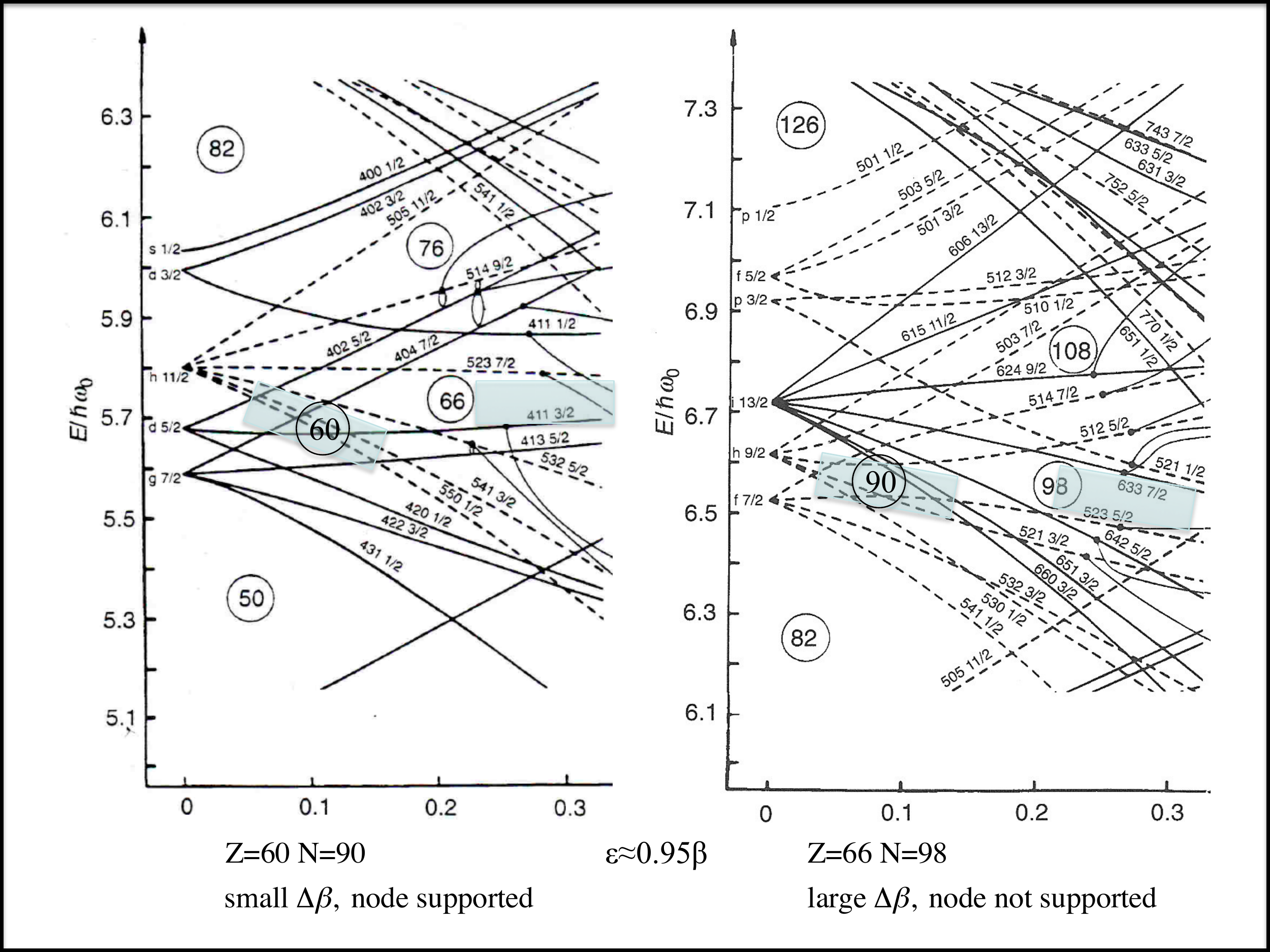}\hspace*{0.04\linewidth}\includegraphics[width=0.48\linewidth]{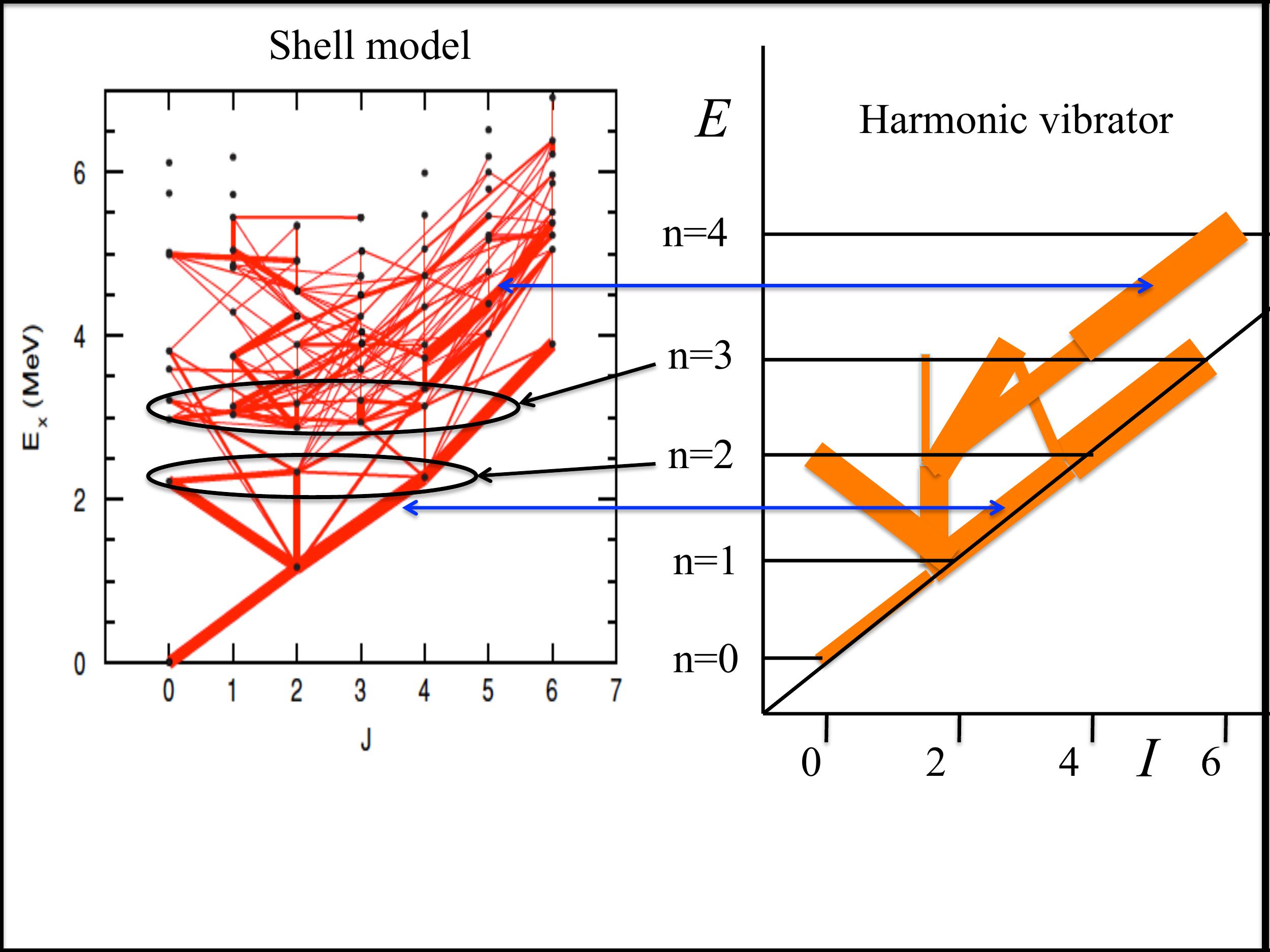}}
\caption{\label{fig:CohBet} Left: Energies of the modified oscillator as function of the deformation parameter $\eps$. 
From Ref. \cite{NR}. Right: Shell Model Calculation for $^{62}$Ni. The width of the bars is proportional to the $B(E2)$ value of the connecting transition. From Ref. \cite{Chakraborty11}. 
The collectively enhanced transitions of the Shell Model calculation are associated with the enhanced transitions  of the harmonic vibrator limit of the BH.} 
\end{figure}

\section{COHERENCE OF THE DEFORMATION DEGREES OF FREEDOM}\label{sec:CL}

The resolution of the collective wave function of deformation degree of freedom $\beta$  is also limited by a coherence length. It appears in the overlap of two 
mean-field solutions with different deformation \mbox{$\vert \langle \beta\vert \beta'\rangle\vert^2\approx\exp \left[-(\beta-\beta')^2/\Delta \beta^2\right]$}.
The overlap plays a central role in describing the shape dynamics by means of the Generator Coordinate Method. To my knowledge, the resolution aspect 
has not yet been paid much attention so far. Here I address it only in a qualitative way.   

The left part of Fig. \ref{fig:CohBet} shows the single particle levels as function of the deformation variable $\beta$. The overlap falls off the stronger the more the occupation of
the states near the Fermi surface changes over an interval of $\beta$. For $Z\approx 60$ and $N\approx 90$  many up-sloping levels cross many down-sloping. This results
in a considerable re-occupation over the indicated $\beta$ interval. A relatively small coherence length $\De \beta$  is expected.  For $Z\approx 66$ and $N\approx 98$ 
there is no re-occupation within the deformation interval. The overlap will still be smaller than one, because the single particle wave functions change with $\beta$. A 
relatively large value $\De \beta$ will result. The difference accounts for the following. In the transitional nuclei around $N=90$ there is
a low-lying  0$_2^+$ state with the properties of the collective one-phonon $\beta$ vibration, but no evidence for the two-phonon state. The coherence length $\De \beta$
is small enough to resolve one node of the vibration but too large to resolve two nodes. For the well deformed nuclei around $N=98$ there is no evidence for a collective 
$\beta$ vibration. The coherence length $\De \beta$ is too large to even resolve one node.

The right part of Fig. \ref{fig:CohBet} shows the E2 transitions obtained in a Shell Model calculation for spherical nucleus  $^{62}$Ni. 
They are compared with the pattern of transition strength of the harmonic vibration limit of the BH. 
As discussed in the Introduction 
(c. f. Fig. \ref{fig:decoherence}) the quadrupole vibrations become increasingly de-coherent when moving way from the yrast line. The Shell Model
gives a collective 2$^+$ state  interpreted as the one phonon state and at twice the energy the states 0$^+$, 2$^+$, 4$^+$ interpreted as the two-phonon triplet.
  As expected for the  harmonic vibrator,  the transition strengths to the one- phonon star are enhanced and there are not transitions to the zero phonon state. 
 In contrast  the $2^+_2\rightarrow2^+_1$ transition is much weaker than the  $4^+_1\rightarrow2^+_1$  and $0^+_2\rightarrow2^+_1$ transitions,
 which are not twice as strong as the  $2^+_1\rightarrow0^+_1$ transition. The transition between the yrast states of the Shell Model are strong and 
 may be accounted for by an anharmonic   tidal wave. The Shell Model calculation shows enhanced transitions parallel to the yrast sequence, which correspond to the 
 rare members of the vibrational multiplets. An extension of the tidal wave approach to describe these states seems promising. The remaining part of the Shell Model
 transition pattern looks chaotic. The coherence of the vibrational motion is lost. 
    
 \section{SUMMARY AND CONCLUSIONS}
 Numerical solutions of the phenomenological Bohr/IBM Hamiltonian with few parameters provide an advanced classification scheme for collective states.
 Experimental $B(E2)$ values for the transitions within the ground state band contradict the finite-boson number assumption of IBM. 
  The Bohr/IBM 
 Hamiltonian derived from various mean-field approaches within the adiabatic approximation reasonably 
 well describes 2$^+_1$, 4$^+_1$, 2$^+_2$ states in even-even nuclei across mass table. 
 The coupling between  the collective and quasiparticle degrees of freedom becomes strong for higher states and in odd-A nuclei.  
 The rotating mean field approach provides a versatile non-adiabatic description of the yrast region. It microscopically describes tidal waves, 
 which represent yrast states of vibrator-like nuclei. The coherence length is introduced, which quantifies the degree of collectivity.  In case of 
 rotation, the coherence angle limits the number of states in a rotational band, which is cut-off by band crossing or band termination. The large
 coherence length for $\beta$ vibrations supports only one node in transitional and no node in well deformed nuclei. 
\\
\\
The work was supported by the U.S. DOE under Contract Nos. DEFG02- 95ER-40934.


\nocite{*}
\bibliographystyle{aipnum-cp}%

\begin{thebibliography}{1}
\bibitem{BMII} A. Bohr and B. Mottelson {\it Nuclear Structure Vol. II. Nuclear
Deformations},  (W. A. Benjamin Inc., London/Amsterdam; Don Mils, Ontario/Sydney/Tokyo; 1975).
\bibitem{Frauendorf15} S. Frauendorf, Int. J.  Mod. Phys. E {\bf 38}, 1541001 (2015). 
\bibitem{Caprio2par} M. A. Caprio {\it Phys. Rev. C} {\bf 68} (2003) 054303.
\bibitem{Zamfir02} N. V. Zamfir {\it et al.}, Phys. Rev. C {\bf 65} (2002) 044325.
\bibitem{Delaroche10} J. P. Delaroche {\it et al.}, Phys.  Rev. C {\bf 81} (2010) 014303.
\bibitem{Garrett97} P. E. Garrett {\it et al.}, Phys. Lett. B {\bf 400}, 250 (19970.
\bibitem{Lesher07} S. R. Lesher  {\it et al.}, Phys. Rev. C {\bf 76}, 034318 (2007).
\bibitem{102PdTidalPRL} A.D. Ayangeakaa {\it et al.}, Phys. Rev. Lett. {\bf 110}  (2013) 102501.
\bibitem{MAFGC14} A. O. Macchiavelli {\it et al.}, Phys. Rev. C 90, 047304 (2014) 
\bibitem{Er168ME2} B. Kotili\'nski {\it et al.}, Phys. Rev. C {\bf 517}, 365 (1990).
\bibitem{RMP} S. Frauendorf,  Rev. Mod. Phys. {\bf 73}, 463 (2001).
\bibitem{WS05}	 Satu\l a, W., Wyss, R. A., Rep. on Progress in Phys. {\bf 68},  131 (2005).
\bibitem{FGS10} S. Frauendorf, Y. Gu, J. Sun, arXiv-id: 0709.0254 (2010).
\bibitem{FGS11}S. Frauendorf, Y. Gu, J. Sun, Int. J. Mod. Phys. E {\bf 20}, 465  (2011).
\bibitem{NR} S. G. Nilsson and I. Ragnarsson, {\it Shapes and Shells in Nuclear Structure}, (Cambridge U. Pub.; 1995).
\bibitem{Chakraborty11} A. Chakroborty {\it et al.}, Phys. Rev. C {\bf 83}, 034316 (2011).
\end{thebibliography}

\end{document}